\documentclass[12pt]{article}
\usepackage{authblk}
\usepackage{geometry}                % See geometry.pdf to learn the layout options. There are lots.
\usepackage{natbib}
\usepackage{float}
\usepackage{graphicx}
\usepackage{amssymb}
\usepackage{epstopdf}
\usepackage{flushend}
\usepackage{url}
\usepackage{multirow}
\usepackage{color}

\def\edit#1{#1}
\def\oldedit#1{#1}

\title{Deception Strategies and Threats for Online Discussions}

\author[a,*]{Onur Varol}
\author[b]{Ismail Uluturk}
\affil[a]{Center for Complex Network Research, Northeastern University}
\affil[b]{University of South Florida}
\affil[*]{Correspondence to: ovarol@northeastern.edu}

\date{Please cite \textit{Varol, Onur, and Ismail Uluturk. ``Deception strategies and threats for online discussions.'' First Monday 23.5 (2018).}}

\begin{document}

\maketitle

\begin{abstract}

Communication plays a major role in social systems. Effective communications, which requires transmission of the messages between individuals without disruptions or noise, can be a powerful tool to deliver intended impact. Language and style of the content can be leveraged to deceive and manipulate recipients. 
These deception and persuasion strategies can be applied to exert power and amass capital in politics and business.
In this work, we provide a modest review of how such deception and persuasion strategies were applied to different communication channels over the years. We provide examples of campaigns that has occurred in different periods over the last 100 years, together with their corresponding dissemination mediums.
In the Internet age, we enjoy access to the vast amount of information and the ability to communicate without borders. However, malicious actors work toward abusing online systems to disseminate disinformation, disrupt communication, and manipulate people by the means of automated tools, such as social bots. 
It is important to study the old practices of persuasion to be able to investigate modern practices and tools. 
Here we provide a discussion of current threats against society while drawing parallels with the historical practices and the recent research efforts on systems of detection and prevention. 
\end{abstract}

\section{Introduction}

Communication is a central part of the society and crucial for human evolution~\citep{kirchner1997evolution}. All forms of living organisms develop or inherit ways to interact with each other~\citep{wiley1983evolution}. Shannon's ground-breaking work formally defines the components of efficient communication systems and introduces concepts about information, noise, and transmission bandwidth~\citep{shannon1949communication}. Through human history, we see many forms of communication: verbal, written, artistic expressions, etc. Even one of the simplest forms of communication, drawing, serves as a tangible record that facilitates communication with future generations. Formation of signals and invention of languages are inevitable for evolving groups and systems to transfer information~\citep{skyrms2010signals}. 
Over the centuries, technology has helped us develop more efficient models of communication. \oldedit{Early use of paper and printing technologies have ensured the longevity of the records.} Invention of the telegraph and the telephone overcame the difficulty of transmitting information over vast distances. These peer-to-peer communication systems mirror our natural interactions. \oldedit{We have also invented different mechanisms to transmit information to larger audiences, instantaneously. The Internet became an archive for \textit{virtually all knowledge} by organizing and storing petabytes of data daily. Recent collections surpass any conventional records of notable events and human narratives in history. Additionally, recent integrated systems have been storing streams of environmental sensory information and mobility of individuals, creating extensive archives that were not possible before.}

\oldedit{Independent from the underlying technology and modes of communication, every communication system consists of three main components: sender, receiver, and the medium for dissemination.} In most cases, transmission between the sender and the receiver is not perfect, and this can be attributed to the interference of noise within the medium or how information is encoded and decoded by the sender and the receiver, respectively.
\oldedit{People have developed strategies that can convey information effectively by mimicking the receiver, adjusting their message, and trying to use different properties of the dissemination medium to be able to bridge the communication gap between two ends.}

\edit{Cybernetics literature describes the systematic processes of meme diffusion. Heylinghen shows the factors that contribute to the success of memes and the process that underlies its spread ~\citep{heylighen1998makes}. He points to a 4-stage process: assimilation, retention, expression and transmission. Note that the first three step are about how individuals adopt, process, and embody new information. Only the last stage describes the process of dissemination, however, fitness of memes relies on sender, receiver, and group properties along with the intrinsic qualities of memes.}

In social psychology, there is a large body of work on persuasion and social influence~\citep{chaiken1996, cialdini1993,wood2000} that talks about various cognitive theories and psychological processes behind how people deceive and influence each other. Guadagno and Cialdini discuss persuasion and compliance in the context of Internet-mediated communications, especially textual messages~\citep{guadagno2005online}.
Examples of language and style matching can be seen in the language mimicry observed in the context of power differentials between discussants~\citep{danescu2012echoes,das2016information,bagrow2017information} and the prediction of message popularity~\citep{tan2014effect}. 

As the information within our reach grows exponentially, attention becomes the limiting factor in its consumption.
Communication between humans is limited due to the evolutionary pressure applied by the finite amount of attention and favors efficiency over clarity~\citep{dunbar1992neocortex}.
Herbert Simon introduced the term \textit{attention economy} to explain human attention as a scarce commodity and the economic theory behind the various information processing strategies~\citep{simon1971designing}.
\edit{Imperfect communication channels, as described by Shannon, are also relevant to human communications. Noise introduced to the communication channel might lead to imperfect transmission or misinterpretation by the receiver.}
We have invented different modes of communication to overcome these attention and noise limitations.  When popularity and influence of the content is an important consideration, information producers should adopt a variety of strategies to convey their messages or use a medium that supports broader dissemination. Large-scale broadcasting of information introduces new channels for information dissemination. Radio, television, and newspapers are some examples of one-to-many communication tools. 

The unprecedented increase in social media use may be the result of our limited attention and desire to reach information fast. \edit{Platforms like Twitter mainly serve as information networks where people follow others to access the content they are sharing. Social connectivity on information networks deviates from offline social network structure, where connections between individuals is less likely to reflect their social relations~\citep{kwak2010twitter}. Facebook and other similar social networks, on the other hand, reflect offline social network structure better where people tend to connect with their friends and colleagues. However, information networks promote formation of connections to maximize our need to access information.} To save time when sharing the same content with larger audiences, we \textit{broadcast}. \oldedit{To acquire relevant information, we \textit{filter} and \textit{prioritize} content to be able process it within our attention span. Long term information storage also addresses the limited attention issue by providing an opportunity to go back and access the necessary information when desired.}
Researchers emphasize the importance of the Internet in the study of mass communication and how theories about communication can be applied to this new medium~\citep{morris1996internet}. \edit{Research by Philip Howard points to the important distinction between traditional and modern campaigning, produced by the shifting political culture~\citep{howard2003digitizing}.}

Every communication system has a certain level of noise and disruptions that impacts the efficiency of the overall system. Temporal durability of the messages and limited attention of the receivers may be some of the more significant challenges for earlier communication systems. Recently, we have been facing more serious problems: deception, censorship, and abuse. \edit{Depending on the platform, malicious actors have several mechanisms available to them that can exploit the social trust between individuals and abuse the limited attention of users. Platforms like Facebook rely on personal connections, and we tend to believe what our friends share and this tendency can be exploited. Information networks like Twitter, where the connections are built by the information need, on the other hand, are more prone to attacks by social bots and misleading content.} 
The volume of available online data enables improved success for manipulation strategies through more accurate micro-targeting models, as social media platforms provide the tools to directly interact with users.
\edit{While each interaction recorded online can be harnessed to develop better systems, adversaries can also use them to test and measure the performance of their malicious strategies just as easily.} Researchers study these problems and develop systems that can prevent manipulation and gaming of the system for power and profit. Efforts to educate Internet users are also very important in the endeavor to prevent the dissemination of unreliable and misleading news. 

In this work, we make a modest attempt to discuss the use of different communication channels and adoption of persuasion strategies. We aim to point out the different facets of deception on communication systems. We focus our attention to highlight the problems that we have been facing in the Internet age.

\section{Interplay between politics and communication}

Politics, in broad terms, can be defined as the process of making decisions that apply to all members of a group governed by the same entity. Alternatively, politics can represent the ideologies of a person who tries to influence the way a country is governed. 
To obtain such power and influence, politicians work towards obtaining trust of citizens and persuading their opposition to influence their attitudes. They are required to have strong communication skills and the ability to use available technologies efficiently to reach their goals.

\begin{figure*}
	\centering
	\includegraphics[width=0.8\columnwidth]{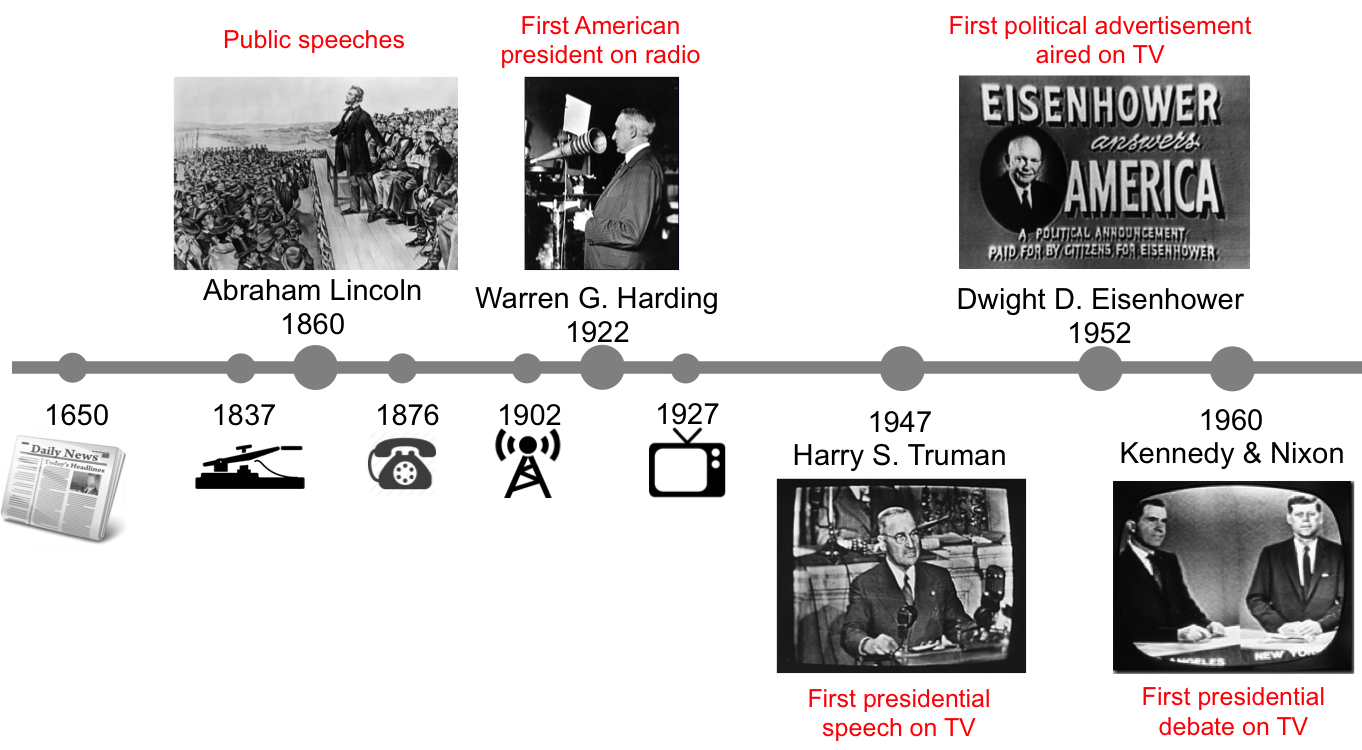}
	\includegraphics[width=0.8\columnwidth]{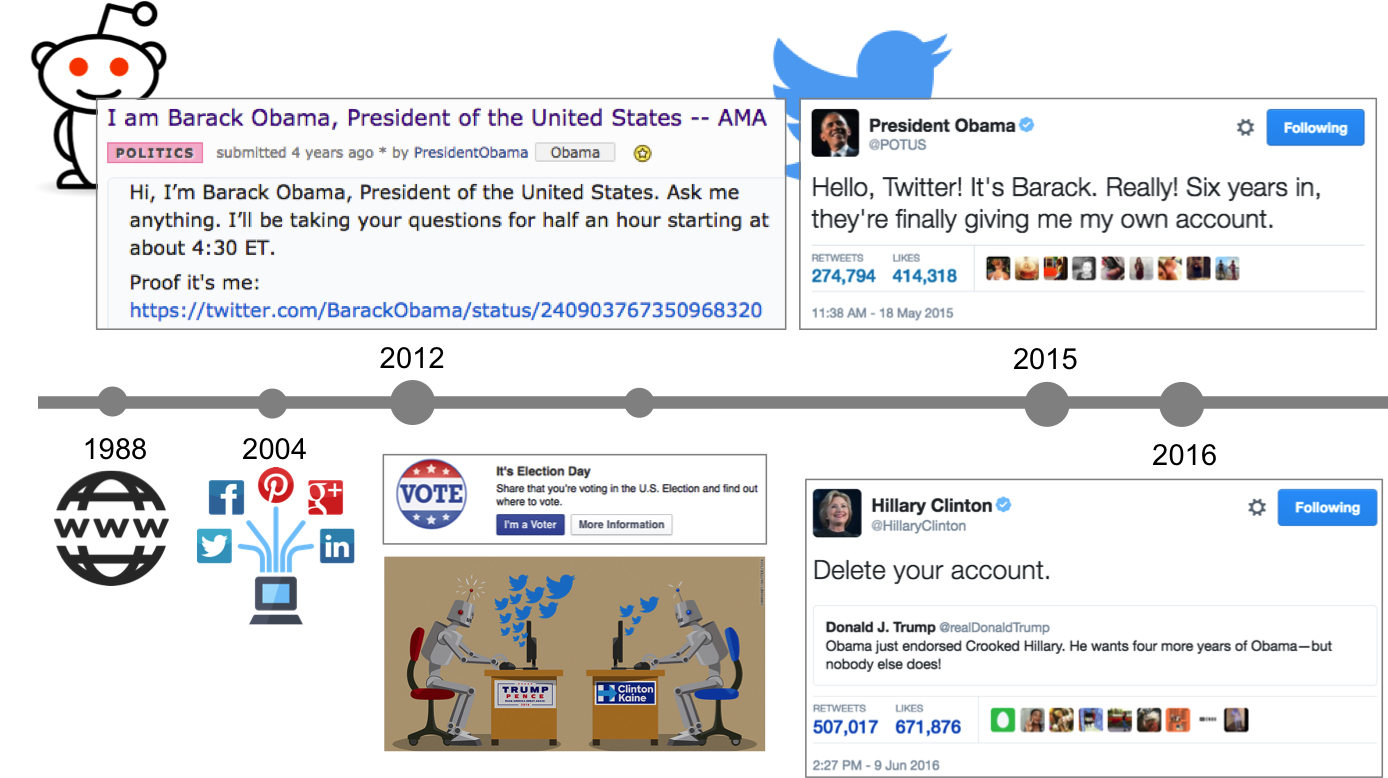}
	\caption{Timeline of US politics and its relation with the technological developments. Some of the leading events are selected. Top panel presents the influence of traditional communication mediums such as newspaper, radio and television. Bottom panel starts with the invention of the Internet and presents examples of US political presence on the internet.}
	\label{fig:politics-timeline}
\end{figure*}

In the political system, we have been observing the impact of different communication media and how politicians adapt their strategies to influence and persuade voters and citizens~\citep{castells2007communication,krueger2006comparison}. We depict a timeline representation of technological development and how politicians adopted these trends in Fig.~\ref{fig:politics-timeline}.

In the early days, newspapers and telegraph were important tools to diffuse news~\citep{even1989number,blondheim1994news}. These technologies accelerated the information diffusion rate from days to hours. \edit{Mobilizing larger crowds for public speeches in parks and squares became easier, thanks to advertisements which made informing a broad audience more convenient.} They also made sharing important policy decisions and public affairs more efficient.

The invention of the telephone and the radio, in 1870s and 1920s respectively,  created opportunities for politicians to reach out to even larger groups. Television, for instance, changed political campaigns significantly~\citep{simon1989impact,behr1985television,west2017air}. 
Only ten years after the first news program aired on the BBC, in 1947, President Truman gave his presidential speech live on TV. This trend was followed by the first TV advertisement placed by Eisenhower in 1952 and the first presidential debate between Kennedy and Nixon in 1960. 
One estimate of President Truman's campaign indicates that he was able to travel more than 31k miles and meet 500k voters in person. Only four years later, Eisenhower was able to reach millions through television advertisements~\citep{diamond1992spot}. It is also important to note that the nature of these political contacts during campaigns have also changed, as studied to learn more about voter behavior and carefully engineered to steer political discourse.

The information age has transformed our experience in various ways. According to an analysis by Pew research center, 65 percent of US adults are actively using social media~\citep{perrin2015social}. The Internet turns out to be a significant resource to study and answer valuable questions about communication in general~\citep{morris1996internet}.
Politicians have also become active users of the social media. They are able to engage with their constituents and campaign on social networks. \edit{Research on online mobilization shows that it is very effective to directly influence political expression, information seeking and real-world voting behaviour of millions of people~\citep{bond201261}.} 

In the last presidential election of the United States, we have observed an active role of social media. 
\oldedit{Researchers have been focusing on analyzing disinformation and social bots to better understand their impact on the election process in many elections around the world, especially the 2016 presidential election of the United States ~\citep{bessi2016social,ferrara2017disinformation,howard2016bots,howard2016bots2}}. We will discuss more on these subjects in the following sections.

\section{Propaganda and campaigns on traditional media}

Traditional communication channels like newspapers, radio, and TV have changed how political campaigns were organized and campaign funds were allocated, so these platforms can be used more efficiently. We provide examples from US politics, however, these observations are applicable to most countries. Here, we will delve into campaign strategies adopted on traditional media channels.

Advertisement has a significant role in reaching voters, and the goal of a successful campaign is to choose the right approach to ultimately win the election. Successful campaigns are often the ones with the most memorable themes and visuals that help sway the public opinion.

Since ancient Greek times, rhetoric and elocution have been recognized as the highest standard for a successful politician. Aristotle's rhetoric describes three main mechanisms for persuasion: ethos, pathos, and logos~\citep{aristotle1992rhetoric}. 
\textit{Ethos} is an appeal to authority, or the credibility of the presenter. If a presenter has credibility and possesses certain moral values, these moral values can be utilized to support a message. Examples of such campaigns were common among cigarette advertisements, showing actors dressed as doctors to mislead audiences.
\textit{Pathos} is another important component, which appeals to the emotions of the audience. Pathos might not only appeal to positive emotions like hope and gratitude, but also to negative emotions such as fear and feeling threatened. Lastly, \textit{logos} is the logical appeal, or the simulation of it. It is commonly used by presenting facts and figures to support claims made by the presenter. It is often used together with ethos. 

Persuasion and propaganda are the main tools in a traditional campaign. All forms of campaign media, such as posters, TV ads., etc., are the products of carefully engineered themes and messages. How public opinion is created and shaped in advertisement campaigns is explained by Edward Bernays in his seminal work ``Crystallizing public opinion'' together with various examples~\citep{bernays2015crystallizing}.

\begin{figure}[!t]
	\centering
	\includegraphics[width=0.255\columnwidth]{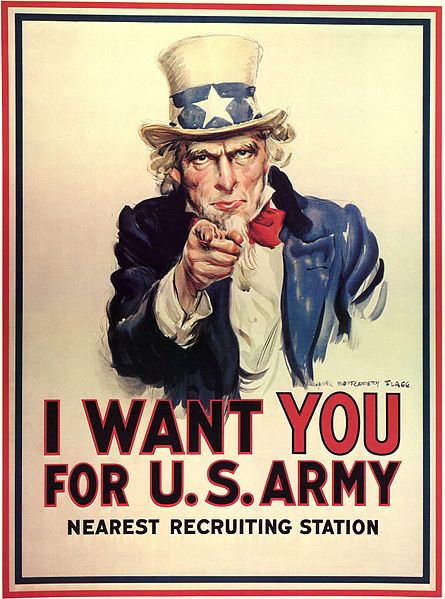}
	\includegraphics[width=0.255\columnwidth]{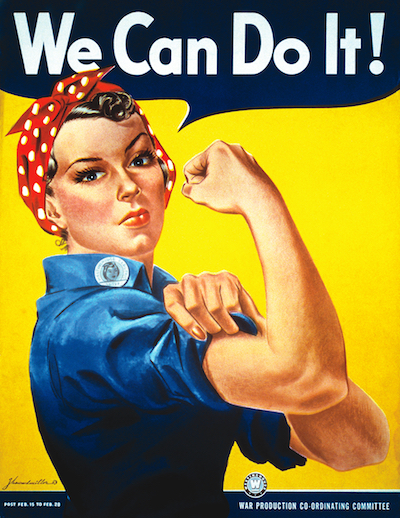}
	\includegraphics[width=0.225\columnwidth]{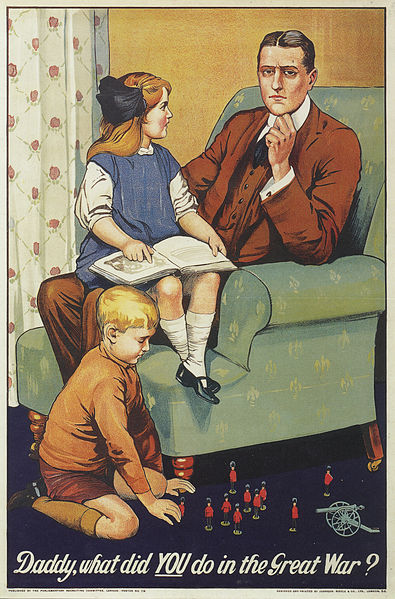}
	\includegraphics[width=0.225\columnwidth]{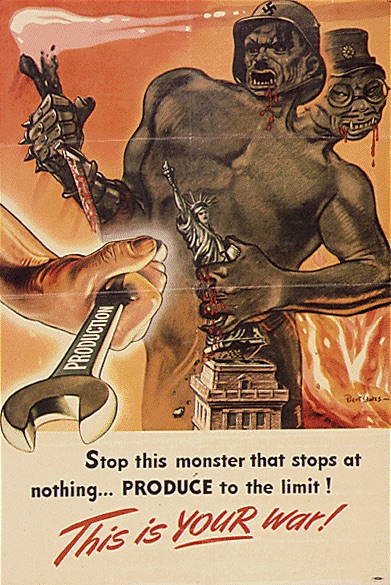}

	\caption{Some of the notable examples of propaganda posters: ``Uncle Sam''~\citep{wikiUncleSam},  ``Daddy, what did YOU do in the Great War?''~\citep{wikiDaddy}, ``We Can Do It!''~\citep{wikiWeCanDoItFem}, and a propaganda poster from the USA against Nazis and Japanese during the WWII~\citep{wikiYourWar}.}
	\label{fig:posters}
\end{figure}

% Propaganda studies
\oldedit{In many cases, persuasion campaigns on politics take the form of propaganda, where persuaders work to achieve a desired response from the targeted audience by following a predefined agenda~\citep{cunningham2002idea,jowett2014propaganda}. Propaganda have been used in the past to recruit people for a cause, manipulate opinions of groups, and create conflicts between two parties.}
Earlier engineered propaganda campaigns used printed media, such as posters and newspaper advertisements, to reach their targeted audiences.

Common themes in propaganda posters are depicting an enemy as evil, or portraying yourself to look righteous~\citep{mahaney2002propaganda}. 
Some of the most memorable posters target various other personal traits and moral foundations as well (see Fig.~\ref{fig:posters}). 
For instance, the ``I Want You'' poster presents Uncle Sam as a way to manifest patriotic emotion, which was used to recruit soldiers for both the first and the second world wars. A similar example of recruitment propaganda was released by the British government during WWI, which depicts a daughter posing a question to her father, ``Daddy, what did YOU do in the Great War?''.
This poster is trying to manipulate an able man with the guilt associated with not volunteering for wartime service.
``We Can Do It!'' is another propaganda poster that was used to encourage involvement of women in the wartime economy during WWII. It later became popular for promotion of feminism and other political issues~\citep{shover1975roles,honey1985creating}. An example of a poster that demonizes the enemy is also presented in Fig.~\ref{fig:posters}. During war, stereotyping of the citizens of hostile countries and their values can often be seen in propaganda posters. 

Perhaps not surprisingly, we observe an increase in comic book sales during international conflicts as well~\citep{murray2000popaganda}. Comic books are used as propaganda tools predominantly by employing visual cues to present cultural ideas embodied in flesh-and-blood characters. Ideas about nationalism, societal stability, and struggles over femininity were best presented by Superman, Batman, and Wonderwoman respectively~\citep{chambliss2012superhero}. \oldedit{These depictions of ideas also create a state of psychological warfare between nations as well. They aim to gain leverage on their opponents without military intervention~\citep{linebarger2015psychological}. The purpose of these propaganda campaigns is to affect the morale of their opponents, while making the current situation more appealing to their own citizens.}

Another interesting example where media helped influence the social psyche is the original Godzilla movie from 1954 \citep{Godzilla1954}. The movie depicts the terrible destruction of Tokyo and its citizens, by an unstoppable radioactive monster whose signature attack is an atomic breath. It has provided the Japanese public, who were the only firsthand witnesses to the terrible powers of atomic weapons, a chance of cathartic relief. It has also allowed wider world audiences to face the devastation that has been brought to Japan, while depicting the nation as an innocent bystander assaulted by forces beyond its control \citep{kalat2017critical}.

Themes and motives used in political television advertisements show common parallels with the propaganda posters used during the Second World War. An analysis of over 800 TV advertising spots between 1960 and 1988 shows that \textit{negativity} in the advertisements is mostly appealing to voters' fears~\citep{kaid1991negative}. 
We observe shared components such as triggering fear and emotions, nationalism, and demonizing the enemy.
Tony Schwartz, a media consultant, has created one of the most memorable election advertisements in US politics. ``Daisy'' spot was aired only once in 1964, but was later replayed several times in other news outlets because of its emotional impact. In this short clip, the association between a countdown for the atomic bomb and a young girl counting daisy petals is utilized to trigger emotional response and fear.

\oldedit{Power of television was widely used by politicians for reaching out to larger crowds, and increasing awareness of selected topics they deemed significant~\citep{diamond1992spot,benoit1999seeing,hermida2010tv,dimitrova2014effects}.}
Advertisements play the important role of putting the ``typical citizen'' on the spot and setting the norms and important questions. Politicians employ advertisements by either supporting their own campaigns, or attacking the policies of their opponents~\citep{simon1989impact,behr1985television}. One of the first examples of this effort was known as the ``Eisenhower answers America'' campaign, where the President answered questions recorded in a studio that contained important messages for his campaign. 

Associating admired celebrities with certain ideologies is another strategy used in political campaigns. 
McAllister discusses the personalization of politicians, and how political priming works through television~\citep{mcallister2007personalization}.

Persuasion is a broad term that covers different types of influence, \oldedit{including deceptive strategies}. We have talked about how advertising is used to influence political beliefs. However, influence through advertisement is not the only type that affects and changes belief systems~\citep{cialdini2001science}. 
Fake news and conspiracy theories are some examples of such deceptive strategies.

Most deception campaigns use strategies that present content along with conflicted facts and distorted claims by authorities~\citep{clarke2002conspiracy,young2010sanctifying}. 
Conspiracy theories are one of the most extreme but persistent examples of disinformation. They appeal to the psychological urge to explain that mysterious things happen for a reason~\citep{goertzel1994belief,sunstein2009conspiracy}. Successful conspiracy theories emerge from a group of supporters, who believe in the sinister aims of higher entities such as governments, religious groups or even extraterrestrial life forms~\citep{goodnight1981conspiracy}.  

\begin{figure}[!t]
	\centering
	\includegraphics[width=0.6\columnwidth]{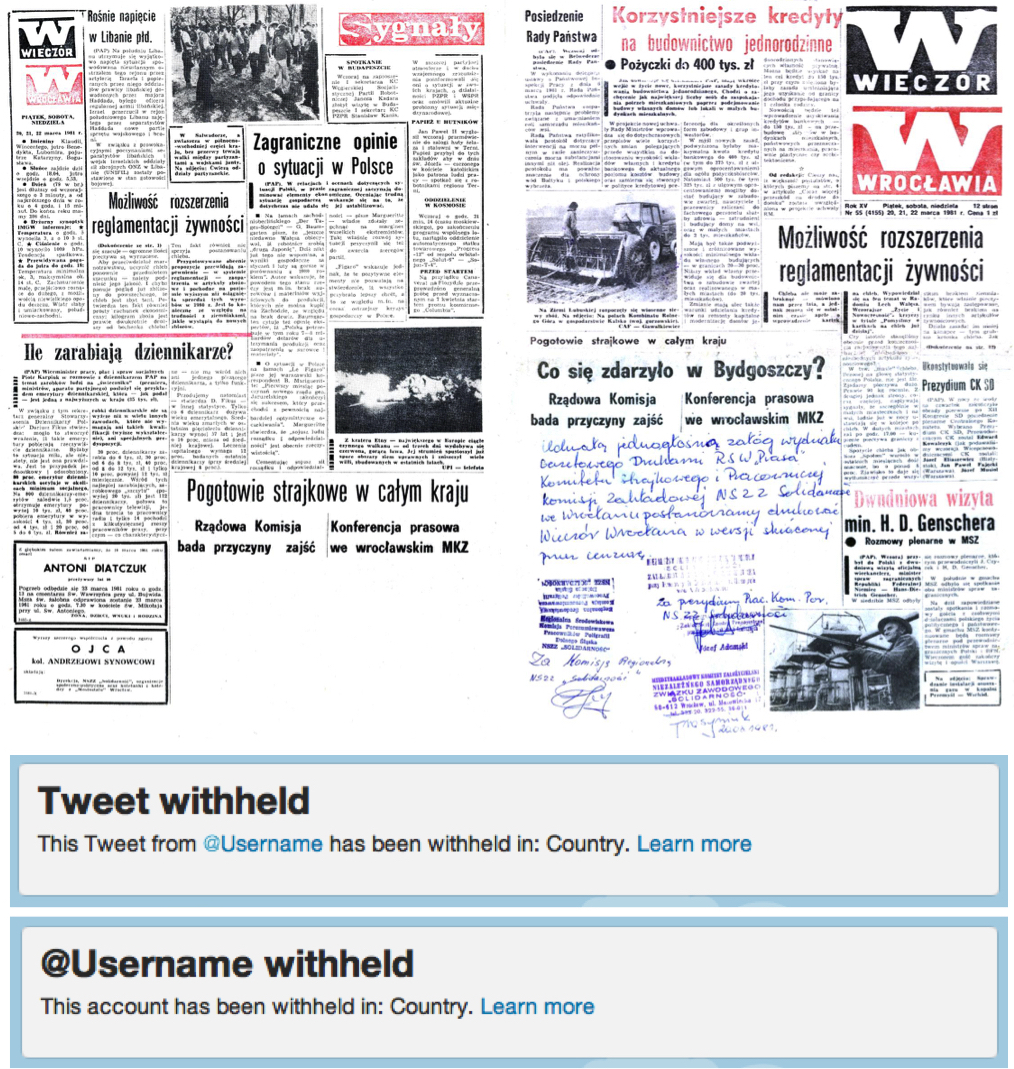}
	\caption{Example of news censorship in Poland~\citep{newsCensorship} on the top. Twitter withheld messages can be seen in the bottom for tweet and user censorship.}
	\label{fig:censorship}
\end{figure}

Censorship is the practice of repressing the dissemination of the truth, or opinions of opposing parties. 
Historically, practices such as collecting printed media, preventing the release of movies, or manipulating pictures or news to hide facts have been observed. 
Nazi Germany and USSR government under Joseph Stalin are both known to have collected books and other printed media and burned them during political repressions~\citep{goldberg2006reading}.
Such practices have inspired dystopian novels like \textit{Fahrenheit 451}~\citep{bradbury2012fahrenheit} and \textit{1984}~\citep{orwell2009nineteen}.

Censorship demands on various newspapers were protested by printing the censored content blank~\citep{collins1996battle}.  
Examples of such counter-censorship tactics can be seen in French, Australian, and Palestinian media.
In Fig.~\ref{fig:censorship}, we present an example of a censored newspaper from Poland dating back to 1981. In response to censorship, the newspaper decided to print the censored section with just the headlines and blank space instead of the text substituted by the censor. 
Recently, Twitter has introduced a similar precaution towards governmental censorship requests with the \textit{withheld tweets} feature. 
Withheld tweets are censored only in the country that made the request to Twitter through official channels like governments and law enforcement. Twitter determines the user locations based on the IP addresses to apply the content censorship selectively. Users accessing Twitter from a censored country are notified by the template tweets in Fig.~\ref{fig:censorship}. 
Therefore, users are notified that the tweets they are trying to access are being censored, serving a similar purpose as the historical examples of white spaces placed by newspaper editors. 

\oldedit{We would also like to note that, censorship can also be applied by disrupting the communication channels and making them nonfunctional for the users, for example, by the excess activity of social bots.} 
An example of such communication disruption through bots was observed in Mexico~\citep{suarez2016influence}.

\section{Deception in the age of the Internet}

Technologically mediated communication systems, like social media platforms and social networks, support the production of information cascades and connect millions of individuals~\citep{goel2015structural,boyd2010social,vespignani2009predicting}. 
Our society has been going through profound changes about 
how we consume and produce information~\citep{aral2012identifying,bond201261}, 
how we interact with our peers~\citep{centola2010spread,centola2011experimental}, and 
how we seek information about societal events surrounding us~\citep{metaxas2013rise,metaxas2012social}.

We are exposed to a tremendous amount of information through social media platforms. Social networks help us organize at least a part of this information, through what our friends share. \oldedit{However, some of the information we share with our network may be inaccurate, and that means we are unintentionally helping the dissemination of misinformation. We may be helping malicious entities, and promoting a disinformation campaign by naively sharing content we find appealing. Our friends in the network are also likely to have similar interests and tendencies as we do, which leads to the spread of misinformation content due to homophily~\citep{mcpherson2001birds}.The echo chambers that we live in also help the amplification of this misinformation~\citep{adamic2005political,conover2011political}.}

\edit{Integrating social media as a dissemination medium provides opportunities for everyone to share their stories and experiences. Some governments take this as a threat to their established system and try to mitigate online activity around certain subjects. One mechanism commonly employed by governments around the world is using Internet censorship, or blocking service providers. These practices terminate the users right to access information.}

Malicious actors on social media can also employ typical misinformation and deception campaigns.
\textit{Astroturf}, for instance, is a peculiar form of deception, often observed on social media in the context of politics and social mobilization \citep{ratkiewicz2011truthy}. It aims to emulate a grassroots conversation through an orchestrated effort. \edit{Although the history of astroturfing and lobbying is older than social networks, these platforms provide a more visible stage for the interactions between online presences of front groups and the promoted accounts~\citep{howard2003digitizing,murphy2012tea}.}
Actors who attempt to generate such orchestrated campaigns generally exploit fake accounts or social bots \citep{hwang2012socialbots,wagner2012social,ferrara2016rise}. 
These artificial means allow the generation of a large volume of content, and emulate the online activity of real users. Cases of massive astroturf campaigns have been observed during political races such as the US senate \citep{mustafaraj2010obscurity} and presidential elections \citep{metaxas2012social,bessi2016social}.

In this section, we have provided examples and mechanisms of campaigns that operate using social bots, disseminate fake news, and apply censorship to restrain the reach of credible information sources. These practices have become very powerful, and the consequences of their involvement remains to be an important research question.

%Twitter: Microphone for the masses? \citep{murthy2011twitter}

\subsection{Social bots}

Increasing evidence suggests that a growing amount of social media content is generated by autonomous entities known as social bots~\citep{varol2017online,howard2017junk,ferrara2016rise,aiello2014people}.
As opposed to social media accounts controlled by humans, bots are controlled by software, algorithmically generating content and establishing interactions. While not all social bots are harmful, there is a growing record of malicious applications of social bots.  Some emulate human behavior to manufacture fake grassroots political support~\citep{ratkiewicz2011detecting}, promote terrorist propaganda and recruitment~\citep{berger2015isis,ferrara2016predicting,bessi2016social,woolley2016automating}, manipulate the stock market or advertisements~\citep{clark2015vaporous,ferrara2015manipulation}, and disseminate rumors and conspiracy theories~\citep{bessi2015science}.  

Discussion of social bot activity, their broader implications on social networks, and the detection of these accounts are becoming central research avenues~\citep{lee2011seven,boshmaf2011socialbot,ferrara2016rise,ferrara2015manipulation}. Magnitude of the problem is underscored by a social bot detection challenge, recently organized by DARPA to study the information dissemination mediated by automated accounts, and to detect malicious activities carried out by these bots~\citep{darpachallenge2015}. 
A recent study on social bots reports that 9 to 15 percent of all active users among English-speaking population exhibits bot-likely behaviors~\citep{varol2017online}. Researchers are also working on identifying social bots with different behavioral patterns and interaction styles. Analysis of social bots during US presidential election points out that more than 20 percent of the accounts were exhibiting social bot behavior~\citep{bessi2016social}. Another analysis points that one third of the total volume of tweets and shared online news articles supporting the political candidates during 2016 election were consisted of fake or extremely biased news~\citep{bovet2018influence}.
Influence of external factors on the US presidential election in 2016 is a source of controversy. Recent research shows evidence supporting the involvement of social bots in political discourse~\citep{bessi2016social}. 

Participation of social bots on political conversations are not necessarily needed to be sophisticated. \edit{Simple interactions such as retweeting can still be powerful, considering the visibility it might lead to. Purchasing fake followers is also common among politicians that wish to create a false impression of popularity~\citep{woolley2016automating}. 
Recent research shows evidence that social bots played a key role in the spread of fake news during the 2016 US presidential election~\citep{shao2017spread} and most of the central actors in the diffusion network consisted of bots~\citep{shao2018anatomy}.}
Social bots are also known to disrupt conversations by flooding a particular conversation channel with content. Pollution of conversations on social media makes it intractable for individuals looking for useful information.

Social bots can also be used for coordinated activities where large collections of social bots, also known as botnets, are controlled by botmasters. Examples of such botnets have been identified for advertisements~\citep{zhou2017starwars} and influencing Syrian civil war~\citep{abokhodair2015dissecting}. \oldedit{The orchestrated behavior of thousands of bots is worrisome, since they can be used to pollute conversation channels, boost the popularity of disinformation content, and target individuals to deceive them on certain subjects.}
Social bots also vary greatly in terms of their behavior, intent, and vulnerabilities, as described by Mitter \textit{et al.}~\citep{mitter2014categorization}.

\subsection{Fake news}

\edit{The term \textit{fake news} is not new~\citep{lippmann2004public}, yet the prevalence of the online disinformation come in different forms and properties: hoaxes, rumors, conspiracy theories, etc.}
Fake news websites deliberately publish hoaxes, propaganda, and disinformation pretending to be a legitimate news source. They often aim to mislead readers, unlike the satirical news websites that may appear similar, in exchange for political and financial gain.

A large amount of disinformation spreads online, and their prevalence and persuasiveness can affect serious decisions around important topics such as vaccination~\citep{kata2012anti,nyhan2013hazards,buttenheim2015mmr}, elections~\citep{allcott2017social,mustafaraj2017fake,giglietto2016fakes,rojecki2016rumors} and stock market behavior~\citep{carvalho2011persistent,lauricella2013twitter} among other issues. A recent study suggests that misinformation is just as likely to go viral as reliable information~\citep{shao2016hoaxy,qiu2017limited}. 
\edit{One of the mechanisms promoting the persistence of fake news is the copycat websites.
Copycat websites operate by duplicating the original content with only trivial changes. If the original article contains misinformation, copycat websites replicate this misinformation as well.} Corrections or removal of the original source are no longer relevant or useful, since many other media outlets are already affected and have already disseminated false information. We can make an analogy between the dissemination of fake news through multiple media outlets and a disease spreading in groups\edit{, vaccination or treatment of a single individual will not stop an epidemic by itself}. \edit{In some cases, even reliable sources can publish misinformation, simply because of the competition introduced by online journalism.\edit{The rush to break the news first has lead news agencies to employ automation tools in their workflows, from tools that write complete articles like the Automated Insights used by Associated Press,\footnote{\url{https://automatedinsights.com/case-studies/associated-press}} to news discovery tools like the Reuters Tracer by Reuters. Reuters Tracer parses through millions of tweets every day and reportedly gives Reuters a 8 to 60 minutes of a headstart on news pieces against its competitors \citep{liu2017reuters}. While automated tools like the Reuters Tracer are being employed to help verify stories in a timely manner as well~\citep{liu2017reuters}, the continuously narrowing window to break a story may lead to articles lacking deliberate investigation which will then be picked up and disseminated by copycat sites before any corrections can be made.} It is almost impossible to propagate corrections to all other copycat articles that are based on the original piece.~\citep{janowitz1975professional}}

Recent research efforts have also focused on modeling the diffusion of misinformation~\citep{del2016spreading,bessi2015science,bessi2015viral,friggeri2014rumor,jin2013epidemiological}. Algorithmic efforts on detecting rumors and misinformation are also crucial to prevent the spread of campaigns with malicious intents~\citep{varol2017early,ferrara2016detection,resnick2014rumorlens,metaxas2015using,qazvinian2011rumor}.

Journalists and readers both have important roles and responsibilities to hinder the dissemination of fake news. Online websites like \textit{FactCheck},\footnote{\url{factcheck.org}} \textit{PolitiFact},\footnote{\url{politifact.com}} and \textit{Snopes}\footnote{\url{snopes.com}} provide fact-checking services to debunk fake news. Fact-checking provided by online services influences the opinions of voters, and provides politicians a guide in judging what news might be fake before disseminating it~\citep{fridkin2015liar,nyhan2015effect}. Researchers are working on designing systems that can evaluate the credibility and truthfulness of claims in order to automate the fact-checking process~\citep{ciampaglia2015computational,wu2014toward}.

The problem with fake news can be partially resolved by educating Internet users. \textit{News literacy} is important, and everyone should at least make an effort to learn how they can identify fake news. 
Recently, we have been observing a growing community of fact-checkers. \textit{Poynter} is one of these organizations, which has released ``International Fact-Checking Network fact-checkers' code of principles''\footnote{\url{poynter.org/fact-checkers-code-of-principles/}} to promote excellence in fact-checking. Another noteworthy example is \textit{First draft}.\footnote{\url{firstdraftnews.com}}
These organizations not only provide fact-checked information about popular claims, but they also monitor political campaigns and elections. 
Collaboration between different fact-checking organizations is promoted by proposing an integrated system to share fact-checking information by implementing \textit{ClaimReview} schema.\footnote{\url{schema.org/ClaimReview}}

\subsection{Censorship}

Preventing censorship and supporting the freedom of speech is crucial in continuing services like social networks, where people can freely express their opinion as long as they avoid disruptive behavior such as abuse and harassment. \edit{According to a report by the watchdog organization Freedom House, out of 65 countries tracked, 49 of them have received a rating of ``Not Free'' or ``Partly Free'' on internet freedom within the observation period of June 2016 to May 2017. This means less than a quarter of the users are living in countries where internet has received ``Free'' designation. While the report also states that the internet is still more free from censorship compared to traditional press, it points out that internet freedom has declined in 32 countries while making mostly minor gains in only 13, showing a downward trend~\citep{House2017}.}

In some societies, governments have responded to the political mobilizations by either terminating the access to online services, or developing laws to restrict the exchange of information~\citep{zhang2006behind}. China, Iran, North Korea, and Turkey are examples of countries applying internet censorship widely. These countries are monitoring social media and news with the intention of controlling the online discourse. If discussions steer into sensitive topics, concerned governments intervene and attempt to control information dissemination~\citep{king2013censorship,ali2013gatekeeping}. 

Platforms like Facebook and Twitter have been censored in the past by limiting the internet access at the country level. 
Social media companies have recently created specialized legal departments to address requests from governments, with hopes to provide continuous service for their users in countries that employ censorship regularly. 
Periodically, transparency reports are released by technology companies like Facebook,\footnote{\url{govtrequests.facebook.com}} Twitter,\footnote{\url{transparency.twitter.com}} Microsoft,\footnote{\url{microsoft.com/en-us/about/corporate-responsibility/reports-hub}} and Google.\footnote{\url{google.com/transparencyreport}} These reports contain the statistics of requests received from different governments, including requests for disclosure of information. The increasing trend in government requests for disclosure of user information and censorship requests are worrisome.  

Many censorship regulations are developed to control or limit dissemination of political discussions. A recent study highlights a significant rate of content removal on Weibo~\citep{bamman2012censorship}. They estimated that 16 percent of the posts compared to overall activity are deleted by authorities on Weibo due to their political content. The content analysis of the censorship on Weibo points out that, content that is aiming to create oppositional awareness towards the Chinese Communist Party are censored more frequently~\citep{vuori2015lexicon}. 
The political impact of micro-blogging platforms was analyzed by comparing Twitter and Weibo use in China~\citep{sullivan2012tale}. 

In another analysis of Weibo, researchers have studied the mechanism of Weibo's trending topic detection system to track sensitive viral discussions~\citep{zhu2012tracking}. Authors have also showed the mechanism behind content filtering by tracking sensitive users~\citep{zhu2013velocity}. They have found that the trend of a sensitive viral topic is short-lived, which points to the effectiveness of Weibo's censorship on sensitive topics.

Twitter requires legal documents to censor content on their platform, unlike the Chinese social media that has centralized control over censorship with no regulatory oversight. 
Twitter announced their ``withheld tweet'' mechanisms to abide by governmental requests in 2012 after the Internet service provider (ISP) level blockages by various governments.
If removal requests are submitted properly by authorized entities, Twitter grants censorship to these requests. Other than content removal, Twitter can limit access to a particular tweet or user when requested by governments. Previous analysis of Twitter withheld content shows that the topical groups that get censored are emerging around politically sensitive topics~\citep{tanash2015known}. There has also been an increasing trend in the amount of censored content on Twitter over the years~\citep{varol2016spatiotemporal}. 

Historically, censorship implies hindering peoples access to content. An alternative mechanism of censorship is the manipulation of the source directly. \edit{A well-known example of such manipulation is the edited photographs from USSR under Joseph Stalin's government, where members of the party that fell out of favor were removed from the pictures. Editorial censorship of newspapers and books are additional examples of such manipulation.} These strategies have become much more difficult in the current online systems, where records of a source are replicated and stored in a distributed manner. However, it is also possible censor content by polluting the communication medium. Finding reliable information within low signal to information ratio channels is becoming a new challenge. It is possible to use social bots to create bursts of posts to distract users and pollute communication channels. An example of such channel disruptions were observed in Mexico recently~\citep{suarez2016influence}, when different hashtags were flooded by social bots, forcing people to move the discussion to alternative channels.

Technical developments like VPN services or the TOR project\footnote{\url{torproject.org}} can provide resilience against censorship attempts. Researchers have also built services to quantitatively measure the censorship problem~\citep{burnett2013making} and analyze the examples of country-wide Internet outages~\citep{dainotti2011analysis,verkamp2012inferring}.

So far, we have briefly described the strategies and mechanisms of deception for traditional media and online platforms. We have also presented recent research efforts for detecting malicious intentions and building platforms to improve the existing situation. 
%In Table~\ref{tab:comparison}, we summarize common properties of problems. 

%\begin{table*}[!t]
%    \centering
%    \begin{tabular}{p{0.1\textwidth}|p{0.4\textwidth}|p{0.4\textwidth}}
%        Property & Traditional media & Social media \\
%        \hline
        
%        \textbf{Medium} & 
%        Television, radio, newspaper and other mass communication media. & 
%        Social networking platform, webpages (news, blogs, etc.), mobile applications \\
%        \hline
%        
%        \textbf{Diffusion} & 
%        Mostly dissemination of physical content or word-to-mouth approach & 
%        Mostly sharing through social media, links in websites, or email \\
%        \hline
%        
%        \textbf{Agents} & 
%        Oppositions groups and fanatics & 
%        Grassroots participants, social bots, trolls \\
%        \hline
        
%        \textbf{Strategies} 
%        & Engineered persuasion campaigns and framing. Yellow journalism to disseminate fake news and propaganda through traditional mass media
%        & Fake news websites (disinformation), orchestrated activities through social bots, and troll attacks (distraction and disruption). Rapid production and evaluation of multiple campaigns possible \\
        %\hline
        
%    \end{tabular}
%    \caption{Comparison between traditional and social media by their major properties.}
%    \label{tab:comparison}
%\end{table*}

\section{Discussion}

We have shown \edit{the mechanisms of} traditional campaigns and modern persuasion techniques so far. Here, we will discuss how we can benefit from the lessons provided by historical evidence and prepare to engage malicious actors proactively. \edit{We will present modern threats for the online echo-chambers and introduce research directions for prevention mechanisms.} 

Current campaigns have been adapting their tactics from the historical examples we have presented to the developments in technology. \oldedit{Tools for dissemination and manipulation have also been evolving and developing alongside campaign tactics to address the new demands of advancements. In terms of human behavior, we are still vulnerable to the similar cognitive biases, which can be used to manipulate opinions of and trigger certain behaviors from individuals and groups. Taking similarities and differences between modern and historical campaigns into consideration, we have the chance to turn the technological and research efforts into our advantage.}

Efforts in designing viral online campaigns have yielded the development of modern marketing tools and strategies. Unfortunately, malicious actors are also able to benefit from such developments and adopt them to achieve their ends. 
Successful campaigns often rely on carefully designed messages and punctual timing. Experts in social psychology can identify possible concepts to frame their campaigns to target specific groups.
Nowadays, the volume and velocity of the data are significantly increased, and thanks to that, evaluating different strategies for manipulation and framing messages have become virtually effortless. 
The abundance of digital data and developments on personalization might result in building targeted campaigns and the ability to rapidly evaluate the effects of different campaign components. \edit{We are living in a data-rich world that helps with accurate estimates of demographics and personal characteristics~\citep{kosinski2013private}.}

\edit{
Researchers demonstrated the predictive power of Facebook data in a 2013 article by predicting various personality traits, demographic information, sexual orientation, and political leaning~\citep{kosinski2013private}. 
Facebook has also published results of their experiment on randomized controlled trial of political mobilization messages during 2010 US congressional elections, in which they delivered messages to 61 Million users~\citep{bond201261}. Facebook demonstrated how minor interventions to the content they deliver can influence real-world voting behavior.
Emotional contagion phenomenon is also demonstrated on Facebook by manipulating the content of the posts~\cite{kramer2014experimental}. This work shows experimental evidence on emotional contagion without direct interaction between individuals.
}

\edit{
The recent example of a controversy about a British political consulting firm called \textit{Cambridge Analytica} demonstrates how a social media platform can become an instrument for political manipulation. The company claims that they have capabilities to build \textit{psychographics} models to predict the propensities of user behavior towards different stimuli such as social media posts with different sentiment and content.
Their data collection methods raise serious concerns about privacy and breach of institutional review board (IRB) protocols. Researchers from Cambridge University have created a Facebook application called \texttt{thisisyourdigitallife} to collect the information of users and their friends for a research project, however this data later got shared with Cambridge Analytica to build models that company has claimed to be effective for micro-targeting.\footnote{niemanlab.org/2018/03/this-is-how-cambridge-analyticas-facebook-targeting-model-really-worked-according-to-the-person-who-built-it} 
Although effectiveness of Cambridge Analytica's methodologies is still not clear, it is important to make our points on data privacy and use of technology for political manipulation.
}

\edit{
Another past case, a UCLA mathematician who has employed \textit{social bots}, \textit{user profiling} and \textit{targeting} on an online dating platform, may serve as a concrete example to help understand how these methods can be employed to achieve real-world results. He has used \textit{social bots}, that were programmed to mimic human users to circumvent safety measures, to mine OKCupid, a popular dating site, for the data from thousands of users. He then used this data to \textit{profile} users and found out that the users fell within one of seven distinct clusters. He analyzed these clusters and decided that he was only interested in users within two out of the seven. Finally, he has used a machine-learning algorithm to \textit{target} these users and help maximize his match percentage with them, resulting in a very large number of users with unusually high match percentages~\citep{poulsen2014math}. It demonstrates that, with the necessary know-how, even a single person with limited resources can achieve meaningful real-world impact.
}

Social bots have been getting increasingly better fake persona generation~\citep{li2016persona,bhatia2017soc2seq} and conversation models thanks to the advancements in deep-learning technologies ~\citep{sordoni2015neural,li2016deep}. Such technologies make detection of social bots more difficult and provide an advantage to the bot creators in this arms race.
Targeted attacks are made possible through the anonymous use of social media, by orchestrating a large army of social bots, trolls~\citep{mccosker2014trolling,aro2016cyberspace}, sock puppets, and bullies~\citep{bellmore2015five,resnik2016celebrities}. 
Examples of extremist activities on social media have been increasing at an alarming rate, and many platforms have started taking precautions for early-detection and prevention of such activities. 
Recent studies have also pointed to the use of social media for the recruitment efforts of terrorist organizations~\citep{berger2015isis,ferrara2016predicting,magdy2015failedrevolutions}. 

Increasing online participation on websites and social media is creating new avenues for and enabling new forms of deception and manipulation~\citep{phillips2018ambivalent}. We have been particularly observing the societal impact of disinformation manufactured with the intent of drawing an extreme reaction, disseminated by cloaked websites and accounts~\citep{daniels2009cloaked,farkas2017cloaked}.  
It is important and valuable to understand the roots of the problem before setting out to propose solutions. 
Confirmation bias is considered one of the contributing  factors~\citep{nickerson1998confirmation}. According to this hypothesis, people tend to believe and seek information supporting their initial opinions. 
An alternative explanation considers the attention paid to judge the credibility of the source. Herbert Simon's work on attention economy might help explain some of our mental shortcuts; we tend to believe a content based on our opinions about the friend who had shared it with us. 
Researchers have studied when readers pay attention to the source of the content~\citep{kang2011source}. They have found that users tend to believe the content based solely on the source they obtained it from, unless the subject is really important to them. Trust towards personal contacts also makes individuals more vulnerable to attacks such as phishing~\citep{jagatic2007social}. These problems can be alleviated by focusing on news literacy. It is possible to restrain the prevalence of fake news when educated online users are combined with appropriate fact-checking tools. A recent review on fake news lays out a future agenda and invites interdisciplinary research efforts to study the spread of fake news and its underlying mechanisms~\citep{lazer2018science}.

Throughout the 20th century, people living in a number of developed countries had access to credible information through accountable sources, where the content was monitored by editorial boards for accuracy prior to publishing. \oldedit{However, in the age of Internet, the mechanisms for information production by journalists and news websites have been changing~\citep{giglietto2016fakes}. These changes are also effecting how we access and consume information. 
Most of the content of an article is created by an original source where many other websites can copy it from, and various services can propagate the content even further. The role of social media in this process is crucial, since users can share those links through their network. These cascades of dissemination can cause problems if the original content has erroneous claims. Correcting the original source after the fact in this process is not likely to fix copycat content and misinformation already disseminated by social media users.}

\oldedit{We should also raise our concerns about the third-party applications connected to social media accounts with user permissions. If these applications are breached by malicious entities, they can be employed to disturb communication channels and disseminate misinformation. A recent example of such an attack have targeted hundreds of Twitter accounts, including popular news organizations and celebrities, when a third-party analytics application was compromised\citep{naziTwitter}.} \edit{These accounts have posted tweets, written in Turkish, that contained the swastika symbol and hashtags which, when translated, mean ``Nazi Germany'' and ``Nazi Holland''. Accounts compromised by this attack, such as Forbes (\texttt{@forbes}), Germany soccer club Borussia Dortmund (\texttt{@BVB}), and Justin Bieber's Japanese account (\texttt{@bieber\_japan}) have the combined follower count of several millions. Considering the wide reach of such a compromise in a 3rd party application is worrisome. }

\edit{Another concern on third-party applications is their ability to change social ties. Applications with necessary permissions can follow and unfollow accounts through APIs. This can lead to long-term manipulation of how people access information by selectively filtering content or providing exposure to certain users. Segregation and filter bubbles are foreseeable threats that can utilize untrustworthy or compromised applications.
Similarly, adversaries can benefit from the changing attitudes towards fake followers to discredit influential accounts by contaminating their followers with social bots. 
It is important to develop methodologies that can be employed by the platforms to identify these attacks, build preventive strategies, and prevent data breaches. 
}

Sir Tim Berners-Lee published a post on the 29th birthday of World Wide Web to share some concerns and challenge our community to make WWW a safer, more accessible, and transparent place for everyone.\footnote{\url{webfoundation.org/2018/03/web-birthday-29}}
There are significant efforts to preserve the social ecosystem. Researchers are developing tools like Botometer\footnote{\url{botometer.iuni.iu.edu}}~\citep{davis2016botornot,varol2017online} to detect social bots on Twitter, 
Hoaxy\footnote{\url{hoaxy.iuni.iu.edu}}~\citep{shao2016hoaxy} to study dissemination of fake news, and
TweetCred\footnote{\url{twitdigest.iiitd.edu.in/TweetCred}}~\citep{gupta2014tweetcred} to evaluate the credibility of tweet contents. The Jigsaw lab of Google also has significant efforts to tackle some of the global security challenges, working on systems and tools to prevent censorship and online harassment.\footnote{\url{jigsaw.google.com}} Considering the impact of technology on the dissemination of misinformation, we share a great responsibility to work together. We should also be aware of the limitations of human-mediated systems as well as algorithmic approaches and employ them wisely and appropriately to tackle the weaknesses of existing communication systems. 
Computer scientists, social scientists, journalists, and other industry partners must collaborate in the effort to implement policies and systems against online threats for an effective resistance.

\footnotesize

\section*{Acknowledgement}
I want to thank our anonymous reviewers, Christine Ogan, and Filippo Menczer
for their insightful discussions and feedback. 

\setcitestyle{authoryear, round, semicolon}
\bibliographystyle{apa}
\bibliography{sigproc}

\end{document}